# XMENTOR: A Rank-Aware Aggregation Approach for Human-Centered Explainable AI in Just-in-Time Software Defect Prediction

Saumendu Roy
Computer Science
University of Saskatchewan, Canada
Saskatoon, Saskatchewan, Canada
plz937@usask.ca

Banani Roy
Computer Science
University of Saskatchewan
Saskatoon, Saskatchewan, Canada
banani.roy@usask.ca

Chanchal Roy
Computer Science
University of Saskatchewan
Saskatoon, Saskatchewan, Canada
chanchal.roy@usask.ca

Richard Bassey
Computer Science
University of Saskatchewan
Saskatoon, Saskatchewan, Canada
apb245@mail.usask.ca

## Abstract

Machine learning (ML)-based defect prediction models can improve software quality. However, their opaque reasoning creates an HCI challenge because developers struggle to trust models they cannot interpret. Explainable AI (XAI) methods such as LIME, SHAP, and BreakDown aim to provide transparency, but when used together, they often produce conflicting explanations that increase confusion, frustration, and cognitive load. To address this usability challenge, we introduce XMENTOR, a human-centered, rank-aware aggregation method implemented as a VS Code plugin. XMENTOR unifies multiple post-hoc explanations into a single, coherent view by applying adaptive thresholding, rank and sign agreement, and fallback strategies to preserve clarity without overwhelming users. In a user study, nearly 90% of the participants preferred aggregated explanations, citing reduced confusion and stronger support for daily tasks of debugging and review of defects. Our findings show how combining explanations and embedding them into developer workflows can enhance interpretability, usability, and trust.

## CCS Concepts

• **Human-Centered Explainable AI**; • **Developer Tools**; • **Software Engineering**; • **Explainable AI**;

## Keywords

LIME, SHAP, BreakDown, XAI Disagreements, User Study

## 1 Introduction

Machine learning (ML) models are increasingly integrated into software engineering (SE) workflows to support tasks such as defect prediction, bug triage, and quality assurance. These models promise to improve software quality and reduce maintenance costs by flagging risky code modules early. However, despite their potential, adoption remains limited because developers often struggle to trust or act on predictions that are presented as opaque explanations rather than intelligible reasoning [19].

XAI methods, such as LIME [39], SHAP [32], and BreakDown [48], attempt to address this by surfacing feature attributions that explain why a model made a specific prediction. However, while these tools promise transparency, in practice, they create new usability challenges that undermine their effectiveness in everyday development. We identify three key usability issues that make current XAI approaches misaligned with developers' needs:

(1) **Conflicting explanations:** Different XAI methods often disagree on which features matter most, how they should be ranked, or whether they contribute positively or negatively. Instead of reducing opacity, this creates cognitive overload, forcing developers to compare multiple explanations without clear guidance on which to trust [28].
(2) **Workflow disruption:** Most explanation tools are delivered through dashboards or external visualizations, detached from the IDE where coding and debugging actually happen [2]. This separation reduces accessibility and slows down fast-paced, context-driven work.
(3) **Erosion of trust and productivity:** When explanations are contradictory, developers report confusion, or outright abandonment of the tools. Over time, this erodes confidence not only in the explanations but also in the predictive models themselves, limiting real-world adoption.

From a human perspective, these issues represent fundamental usability breakdowns. Effective explanations should align with the mental models of developers [29], reduce cognitive load [7], and foster trust and usability [1]. However, current XAI approaches too often do the opposite. They overwhelm users, disrupt workflows, and obscure rather than clarify. To address these problems, we propose a human-centered aggregation approach that treats explanation not as an isolated technical output but as an interaction design challenge. Our method, XMENTOR, reconciles multiple explainers into a single, consistent explanation by applying rank-aware aggregation with fallback strategies for coverage. Crucially, XMENTOR is implemented as a VS Code plugin, embedding explanations directly into the developer's workflow through panels, highlights, and tooltips. This integration reduces the burden of cross-comparison, provides just-in-time feedback where code is written, and allows developers to act on explanations without leaving their environment. In this paper, we contribute:

(1) An **aggregation method** for combining the outputs of multiple XAI techniques (LIME, SHAP, BreakDown) to reduce conflicting explanations.
(2) An **IDE plugin** that integrates aggregated explanations directly into the developer workflow, making XAI more accessible and actionable.
(3) A comparative **user study** evaluating the developer experience with and without the aggregation method, highlighting the effects on comprehension, trust, and usability.
(4) **Design implications** for human-centered XAI tools, emphasizing how explanations can be both reliable and accessible in real-world software development contexts.

These contributions motivate the following research questions:

**RQ1. To what extent do popular post-hoc explanation methods (LIME, SHAP, BreakDown) disagree when applied to defect prediction tasks?** Different explainers often conflict, but we lack evidence of how severe this is in defect prediction [28, 47]. We expect to provide a clear measurement of disagreement (feature, rank, sign) across methods.

**RQ2. Which type of disagreement, ranking inconsistencies or sign mismatches, is more prevalent among LIME, SHAP, and BreakDown?** Understanding whether disagreements stem from feature selection, ranking, or sign mismatches the most critical challenges [1]. Prior work suggests that explainers often agree on contribution direction but diverge on feature importance, making ranking conflicts more prevalent and practically disruptive because developers rely on rankings to prioritize inspection.

**RQ3. Can a rank-aware aggregation method generate more consistent explanations by reconciling these disagreements?** Existing aggregation approaches are too simplistic, we need a structured way to unify outputs [15, 32]. We expect that a rank-aware method will produce explanations that are more stable and coherent.

**RQ4. How do practitioners perceive the usefulness and clarity of aggregated explanations, particularly when integrated into a familiar development environment such as VS Code?** Explanations must feel usable and actionable in real developer workflows [7, 29, 50]. We hope that practitioners will find aggregated explanations clearer, more trustworthy, and easier to use.

By reframing explanation disagreements as an HCI problem, this paper advances the design of human-centered explainability tools that not only open the "black box" but also deliver explanations that are usable, trustworthy, and actionable in practice.

**Replication package** that includes the scripts and data to answer our RQs can be found in our online appendix [3].

## 2 Related Work

Explainable AI (XAI) has been widely studied in domains where transparency and accountability are critical, such as healthcare and finance, as well as in HCI research on trust and fairness [6, 10, 14, 16]. Similar challenges around clarity, completeness, and alignment with human reasoning also arise in software engineering, where explanations must additionally fit naturally into developers' workflows for tasks like defect prediction and debugging [20, 41].

### 2.1 Explainable AI and Disagreement in Software Engineering

As machine learning becomes increasingly common in software engineering tasks such as defect prediction and code review, explainable AI (XAI) has emerged as a key mechanism for making model decisions understandable to developers. Post-hoc explanation methods including LIME [39], SHAP [32], and BreakDown [48] are widely used to attribute predictions to software metrics such as code churn, ownership, and complexity [19, 54]. While these techniques offer valuable insights into black-box models, they are built on different assumptions and approximation strategies, which often leads them to highlight different features or even assign opposite contributions for the same prediction [28, 42].

Recent studies have shown that such disagreements are not merely technical inconsistencies but have direct implications for developers. Conflicting explanations can increase cognitive load, make it harder to decide which factors to act on, and ultimately undermine trust in defect prediction tools [1, 47]. Prior human-centered research further suggests that explanations must be coherent, intelligible, and actionable to support effective decision-making, as overly complex or inconsistent outputs reduce users' confidence and willingness to rely on AI systems [7, 29]. Together, these findings highlight the need for approaches that explicitly address disagreement across explainers while remaining aligned with developers' workflows and reasoning processes.

### 2.2 Human Factors in XAI

A growing body of work has emphasized that explainable AI is not just a technical challenge but also a human-centered one, where the usefulness of explanations depends on how people interpret them. Research in HCI has shown that explanations can increase trust in AI systems but may also introduce cognitive overload and uncertainty if they are inconsistent or too complex. For example, Abdul et al. [1] mapped the landscape of intelligible systems and argued that explanations must be designed for interpretability and accountability rather than technical fidelity alone. Similarly, Kulesza et al. [29] introduced the concept of "explanatory debugging," showing that inconsistent or poorly aligned explanations increase mental effort and reduce user confidence. Cai et al. [7] further demonstrated that explanation design directly impacts how much users trust and rely on AI recommendations, while Cheng et al. [13][4] observed that developers struggle when faced with multiple, conflicting explanation outputs. These findings highlight that overload and uncertainty are major obstacles to adopting XAI in practice, reinforcing the need for aggregation approaches that present a coherent narrative rather than fragmented outputs.

## 2.3 IDE-Integrated Tools

Within SE, IDE-integrated tools have long been developed to support tasks such as testing assistance, automated refactoring, and debugging. For instance, Murphy-Hill and Black [33] studied refactoring tools in Eclipse and found that integration into the IDE is critical for adoption, while Perscheid et al. [36] showed how debugger integration can improve developers' reasoning about program state. More recently, Amershi et al. [2] offered guidelines for human-AI interaction, emphasizing that AI features should be embedded into the user's natural workflow to reduce friction. However, despite this progress, little work has focused on integrating XAI into IDEs in ways that address practitioner needs. While recent work emphasizes embedding AI features into natural workflows, little effort has focused on integrating explainable AI (XAI) into IDEs. Existing tools often present explanations externally, disrupting the developer's workflow. This work bridges that gap by embedding a rank-aware aggregation approach into VS Code, making explanations accessible, trustworthy, and seamlessly integrated into developers' existing environments.

## 2.4 XAI in Foundation Models

Many defect prediction pipelines deployed in practice remain based on structured software metrics and classical ML models. Importantly, the usability challenges we study, like conflicting explanations, cognitive overload, and poor workflow integration, are not unique to tabular models, but arise whenever developers must reconcile multiple explanatory signals [43]. In FM-based workflows, such as defect prediction with code language models or LLM-assisted code review [26], explanations may originate from multiple sources, including token-level attributions, exemplar-based evidence, or model-generated rationales [52]. These explanation channels can diverge in feature emphasis, prioritization, or directional interpretation, creating similar forms of disagreement as those observed with LIME, SHAP, and BreakDown.

XMENTOR contributes a disagreement-aware aggregation and IDE-embedded interaction pattern that is model-agnostic and thus complementary to FM-based approaches. While our current instantiation uses a tabular predictor to isolate and study explainer disagreement, the aggregation logic and human-centered design principles are directly transferable to FM-based defect prediction by aggregating token-level attributions or rationale evidence. We therefore view XMENTOR as an interaction-level contribution that addresses explainability challenges common to both classical ML and foundation-model-based software engineering systems.

# 3 Methodology

Figure 1 summarizes our methodology. We train a defect prediction model, generate explanations using LIME, SHAP, and BreakDown, analyze their disagreements, and apply an aggregation approach implemented in XMENTOR. Finally, we evaluate the aggregated explanations through practitioner feedback.

## 3.1 Data Preparation and Feature Selection

We collected a defect prediction dataset (e.g., ActiveMQ, Camel, Derby, Groovy, HBase, Hive, JRuby, Lucene, Wicket) from a publicly available corpus of Java projects [53]. The dataset includes various software metrics, such as static code metrics, process metrics, and ownership metrics extracted from Git. Files are labeled as `Clean` or `Defect` based on whether they were affected by issue-fixing commits in subsequent releases. We employed automatic and manual feature selection methods to identify key metrics for defect prediction. Automatic selection was performed using the AutoSpearman method [23], as used in prior studies [22, 37, 38]. Alongside automated methods producing 20–30 features, we manually selected six key features using domain knowledge in software engineering and defect prediction. These features, like lines of code and number of changes, offer a more interpretable and developer-friendly alternative to larger automated sets. To address class imbalance, we apply SMOTE [11, 46], and we log-transform skewed features (e.g., `LinesOfCode`) to stabilize distributions.

## 3.2 Model Selection

Prior studies recommend supervised ML models such as logistic regression, random forest, gradient boosting, and neural networks for defect prediction [8]. We adopt Gradient Boosting due to its strong performance on tabular software metrics and its widespread success in defect prediction benchmarks [17, 30]. It provides a good balance between accuracy and interpretability, making it well suited for post-hoc explainability with methods like LIME. The trained model is integrated into the XMENTOR VS Code plugin, where predictions from multiple explainers are aggregated and presented directly within the IDE to support developer workflows.

## 3.3 Aggregation Method

To help developers interpret conflicting explanations, we designed a disagreement-aware aggregation algorithm and integrated it into our IDE plugin, XMENTOR. The algorithm merges LIME, SHAP, and BreakDown into a single unified explanation (Figure 2) by explicitly addressing feature, rank, and sign disagreements through thresholding, rank agreement, and sign agreement, with fallback mechanisms to maintain clarity and interpretability.

**Threshold Setting:** The first step is to decide how many features should appear in the final explanation. We adaptively set this threshold (k) based on the total number of features (n): three if the feature space is small, five if moderate, and ten if large. This rule keeps explanations concise and interpretable while scaling naturally with dataset complexity.

**Rank Agreement:** Next, the algorithm reconciles differences in feature ordering:

(1) In strict mode, the most common feature is selected at each rank in all explainers, while inconsistent alternatives are discarded. If there is no agreement, one feature is chosen to proceed.
(2) In the looser mode, applied when strict filtering produces too few features, the algorithm iteratively collects the most common first-ranked features until the threshold $k$ is reached.

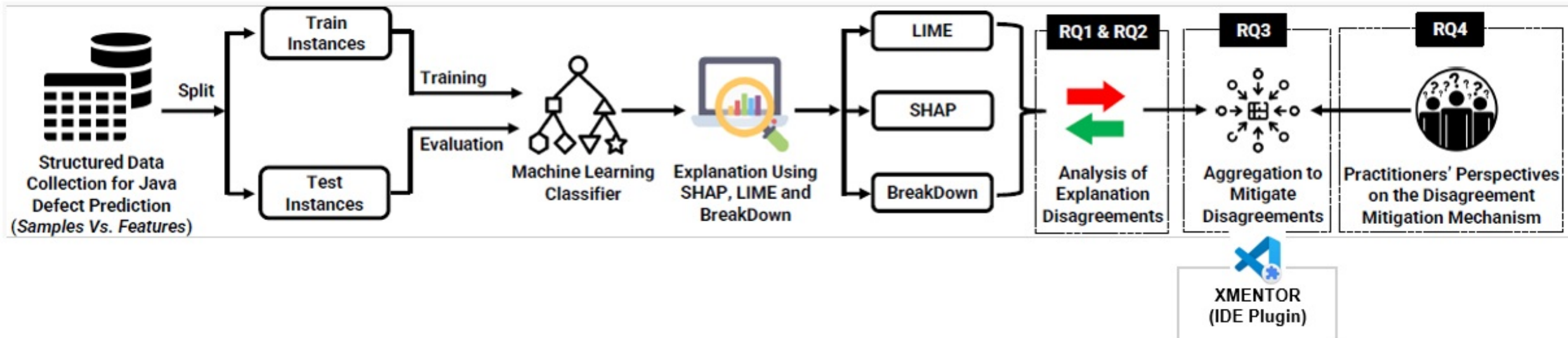

**Figure 1: Schematic diagram of our study.**

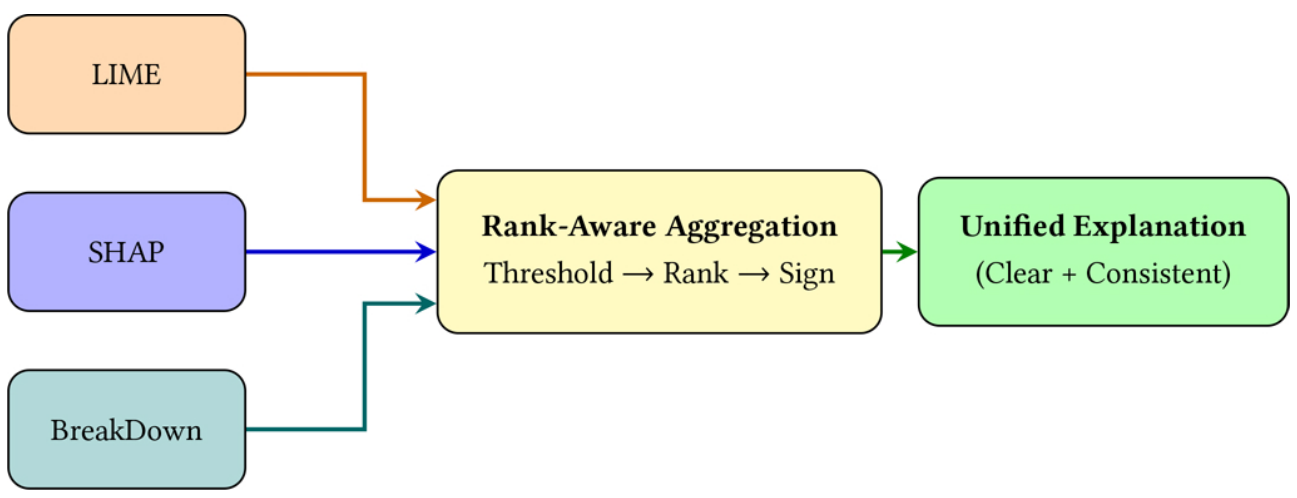

**Figure 2:** **Rank-aware aggregation and unified explanation process.**

This step ensures that features consistently emphasized across explainers are prioritized, helping developers focus on the most stable signals.

**Sign Agreement:** Once the ranks are aligned, the algorithm enforces consistency in the directional contribution:

(1) In the strict mode, a feature is retained only if all explainers agree on its sign (positive, negative, or neutral).
(2) In the looser mode, applied when the explanation becomes too sparse, a feature is kept if a majority of explainers agree, even if one disagrees.

This guarantees that the final explanation avoids contradictory signals that might confuse developers. Finally, if more than $k$ features remain, only the top $k$ are shown. If fewer remain, the looser rules ensure the list is expanded to meet the threshold. This balance prevents explanations from being either too overwhelming or too thin to be useful. For example, we compare three explainers (LIME, SHAP, BreakDown) on seven features ($F1 \dots F7$). Their rankings and contribution signs are shown in Table 1.

**Table 1: Rankings and contribution signs across explainers.**

| Explanation | Rank 1 | Rank 2 | Rank 3 | Rank 4 | Rank 5 | Rank 6 | Rank 7 |
|---|---|---|---|---|---|---|---|
| LIME | F1 (+) | F2 (-) | F3 (+) | F5 (-) | F4 (+) | F6 (0) | F7 (0) |
| SHAP | F1 (+) | F3 (-) | F2 (+) | F6 (+) | F5 (-) | F7 (0) | F4 (0) |
| BreakDown | F1 (+) | F3 (+) | F2 (-) | F5 (-) | F4 (-) | F6 (0) | F7 (0) |

With threshold $k = f(7) = 5$:

- **Strict Rank Agreement:** $F1, F3, F5, F4$ are selected → only 4 features ($< k$).
- **Looser Rank Agreement:** Add $F2, F6, F7$ for coverage.
- **Strict Sign Agreement:** Only $\{F1, F5, F7\}$ remain → too few.
- **Looser Sign Agreement:** Majority signs restore $F3, F2, F6$, giving 6 features.
- **Final Adjustment:** Keep the top 5 since $6 > k$.

**Final Aggregated Explanation:** $\{F1, F3, F2, F5, F6\}$

This example shows how strict rules enforce agreement, while fallback rules guarantee enough features to keep explanations useful and interpretable. Instead of showing three competing explanations (LIME, SHAP, BreakDown), we aggregate them into a unified signal that highlights consensus while reducing noise, as shown in Figure 11.

## 3.4 IDE Plugin: XMENTOR (eXplainability disagreeMENT mitigaTOR)

Human-centered XAI research emphasizes that explanations should be delivered within the developer's workflow rather than through external dashboards [2]. Embedding explainability directly into the IDE reduces friction and makes explanations available at the moment of code review or debugging. Accordingly, our VS Code plugin (Figure 3) analyzes commit changes, extracts 26 software metrics, predicts defect risk, and generates explanations using LIME, SHAP, and BreakDown, which are then combined using a rank-aware aggregation algorithm to resolve disagreements across explainers.

Our plugin, XMENTOR, integrates the rank-aware aggregation algorithm into VS Code, one of the most widely used IDEs. It provides three core interface features:

(1) **Explanation Panel:** A dedicated panel displays the aggregated explanation as a ranked list or bar chart. Developers can toggle between the aggregated view and individual explanations (LIME, SHAP, BreakDown) for transparency.
(2) **Inline Highlights:** Key features influencing defect predictions are visually highlighted in the code editor, helping developers quickly connect explanation outputs with specific code elements.
(3) **Tooltips and Detail-on-Demand:** Hovering over a highlighted feature reveals additional context, such as whether the contribution is positive or negative and how consistent it is across explainers.

These design choices reduce cognitive load and help developers make informed decisions directly in the IDE. Figure 4 contrasts non-aggregated explanations, which require manual reconciliation of multiple outputs, with XMENTOR's aggregated view, which presents a single, streamlined explanation panel. This comparison shows how integration and aggregation turn explainability into a coherent, in-context part of the development workflow.

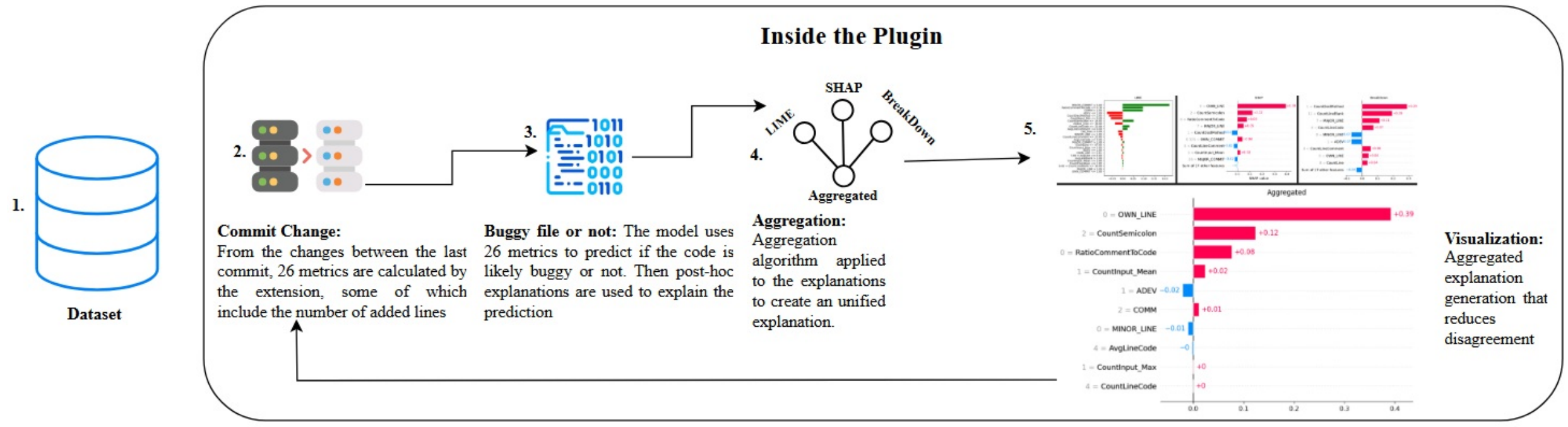

**Figure 3: Workflow inside the VS Code plugin**

# 4 Empirical Evaluation

## 4.1 Quantitative analysis of disagreements

Defect prediction tools must explain their reasoning clearly; otherwise, practitioners may treat predictions as false positives and abandon the tool [31]. As a result, XAI techniques are increasingly used in defect prediction [22, 37], and prior work shows that 82% of developers and managers find discussing defect predictions valuable [24]. Post-hoc methods such as LIME, SHAP, and BreakDown explain predictions through feature importance scores, indicating whether features push a prediction toward or away from a defect [32, 39, 48]. Extensions like TimeLIME [35] and PyExplainer [37] further improve contextual relevance and actionability. Although intrinsically interpretable models, such as attention-based neural networks, are also explored [21], this paper focuses on LIME, SHAP, and BreakDown because they are widely adopted, well cited, and supported by mature open-source APIs [4].

One of the main challenges in XAI is the lack of a definitive ground truth for determining which post-hoc explanation is correct. Because of this, data scientists often rely on multiple XAI tools to interpret the same model decision [28]. However, explanations from different tools often disagree, as shown in Figure 5, which highlights three types of disagreements in feature importance:

- The top-3 features are often not be the same.
- Even if the top-3 features are the same, their importance rankings may differ.
- Even if the features and rankings match, the signs (positive or negative impact) of their importance may differ.

While the example uses the top-3 features, disagreements can also occur for the top-k features, depending on the number of features in the model. Krishna et al. [28] formalized these disagreements with three metrics: feature agreement (FA), rank agreement (RA), and sign agreement (SA).

Figure 5, and Table 2 show how LIME and SHAP provide conflicting explanations, making it difficult for practitioners to gain meaningful insights. This inconsistency raises concerns that users may lose trust in these tools and stop using them altogether. Addressing the disagreement problem in a SE context is therefore essential. Previous studies, like Shin et al. [44], examined how data sampling and classifier choices impact post-hoc explanations, while others focused on developing more robust methods to handle such inconsistencies [12, 40, 51].

**Table 2: Comparison of feature importance between LIME and SHAP, highlighting differences in rank, weight, and sign.**

| Feature | LIME Rank / Weight | SHAP Rank / Value | Agreement |
|---|---|---|---|
| CountLine | Rank 1 / 0.24 | Rank 1 / +0.22 | High (rank consistent) |
| CountPath_Max | Rank 2 / 0.09 | Rank 3 / +0.08 | Partial (different order) |
| CountPath_Mean | Rank 3 / 0.07 | Rank 4 / −0.03 | Disagree (sign + rank) |
| Added_lines | Rank 4 / 0.06 | Rank 5 / +0.01 | Disagree (sign + rank) |
| Del_lines | Rank 5 / 0.06 | Rank 2 / +0.14 | Disagree (rank mismatch) |

This analysis quantified explainer alignment and divergence. Results showed frequent mismatches between feature and sign, with rank disagreements most common, highlighting the need for rank-sensitive aggregation.

## 4.2 User Study

To complement the quantitative findings, we conducted a user study with 42 software practitioners to evaluate the perceived clarity, usefulness, and trustworthiness of aggregated explanations, as well as the performance of the XMENTOR tool. While the quantitative analysis revealed where and how disagreements occur among LIME, SHAP, and BreakDown, it could not capture how practitioners experience these disagreements or whether aggregation meaningfully improves their ability to interpret and act on predictions. The user study therefore followed a mixed-method design with both structured tasks and survey responses, allowing us to assess not only technical outcomes but also human-centered perceptions of clarity, trust, and practical utility.

We surveyed practitioners to validate the proposed aggregation approach and gather their preferred alternatives. The survey followed the personal opinion guidelines of Kitchenham and Pfleeger [27]. We also consider the code of ethics [45]. Notably, the survey was thoroughly revised and approved by the institutional Behavioural Research Ethics Board.

**Survey Design:** Our survey, which contains both multiple-choice and free-text questions, begins by outlining its purpose and research goals, stressing the confidentiality of participants' responses. The study was structured in two parts. The first part

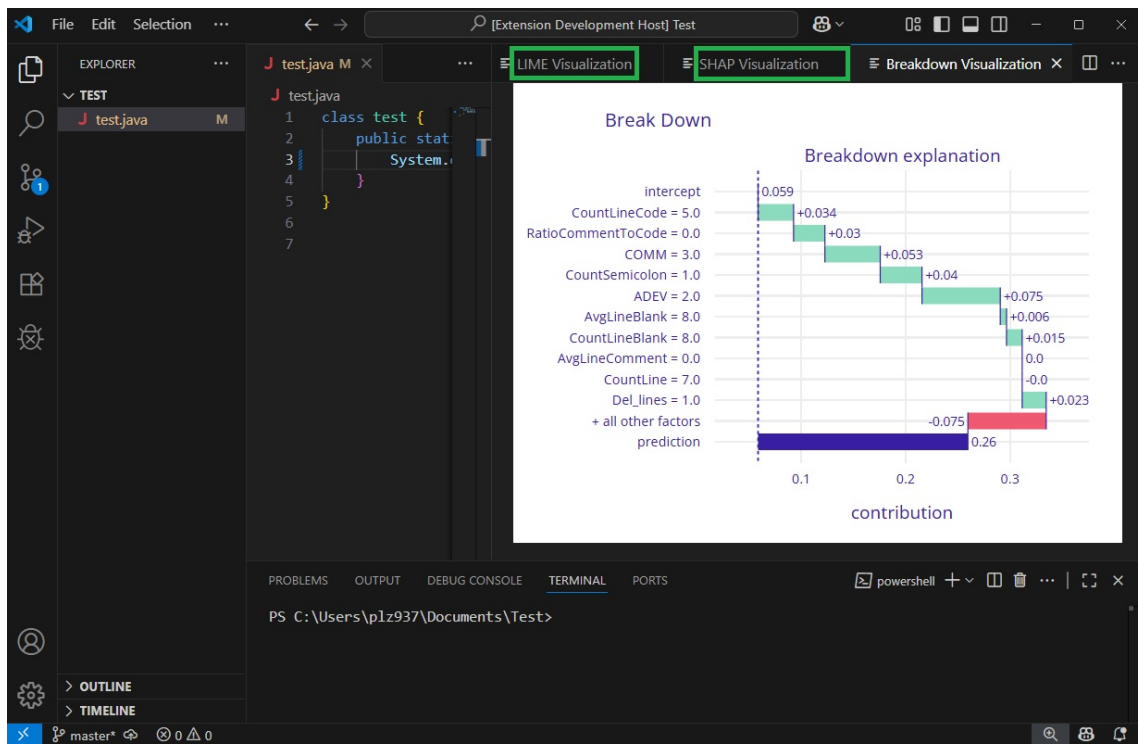

(a) IDE view without aggregation: developers must compare LIME, SHAP, and BreakDown separately.

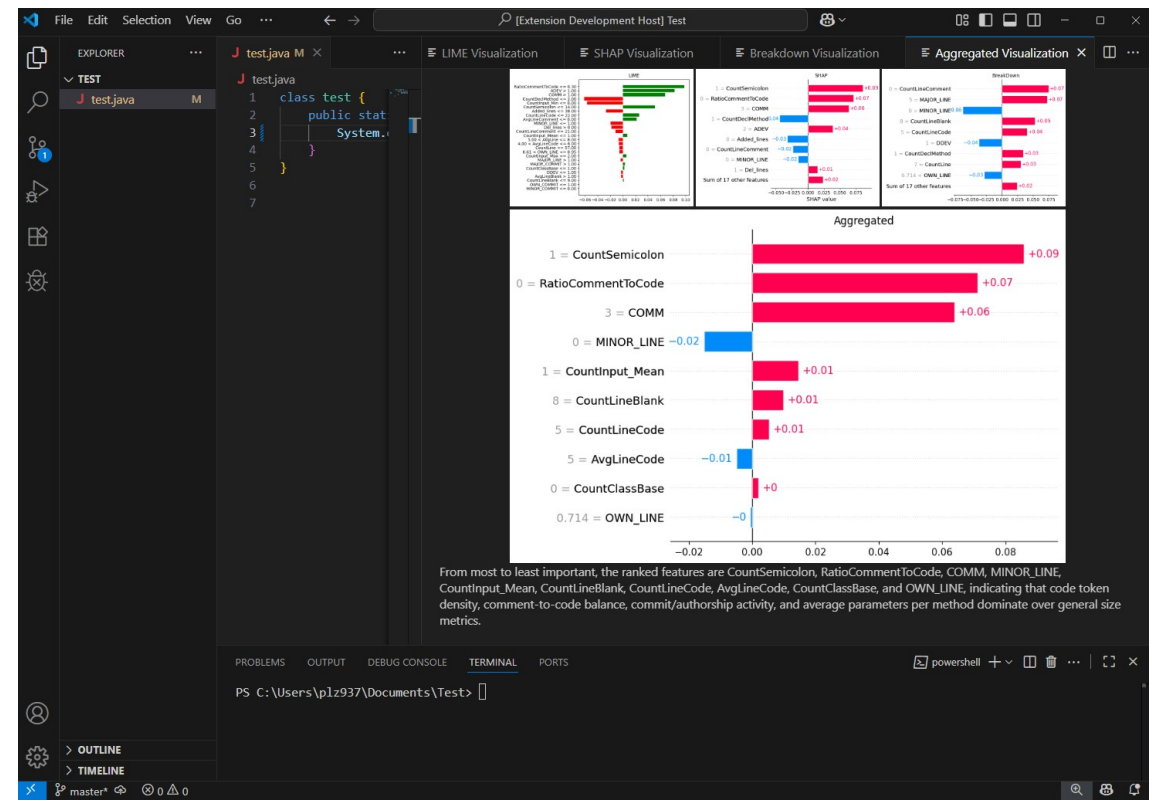

(b) IDE view with aggregation: XMENTOR presents a single, unified explanation.

**Figure 4: IDE integration comparison: Without aggregation, explanations from LIME, SHAP, and BreakDown appear fragmented. With aggregation, XMENTOR delivers a single coherent view in VS Code, reducing cognitive load and improving clarity.**

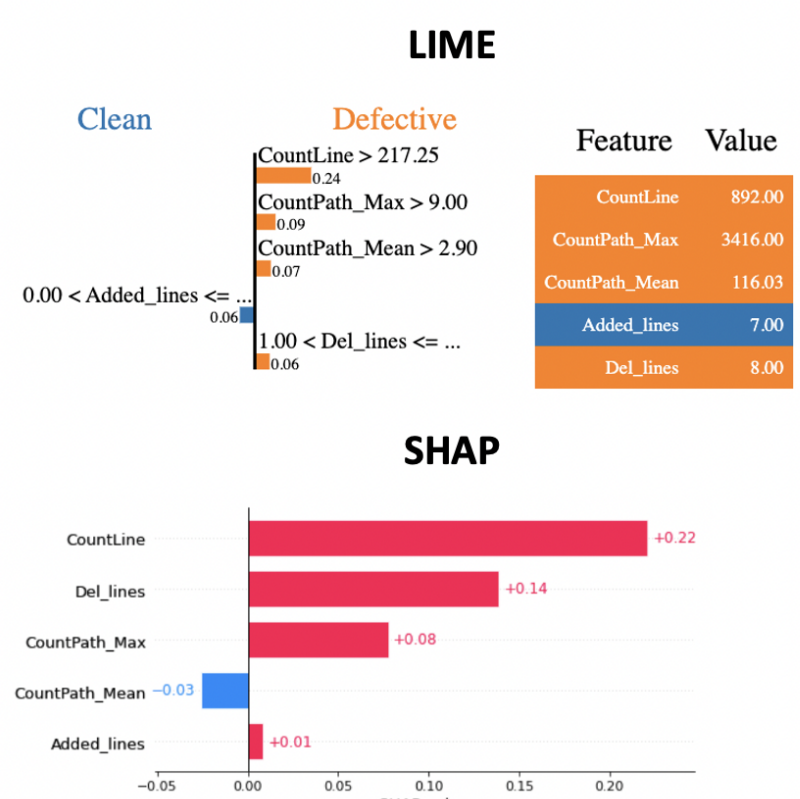

**Figure 5: Example of a LIME and SHAP explanation.**

**Table 3: Experience and Profession of Participants (MLE/C: Machine Learning Engineer/Consultant, AIR: AI Researcher, DAS: Data Analyst/Scientist, SE: Software Engineer, AC: Academician)**

| Experience (Years) | | | | Profession | | | | |
|---|---|---|---|---|---|---|---|---|
| ≤2 | 3–5 | 6–10 | ≥10 | MLE/C | AIR | DAS | SE | AC |
| 7 (16.7%) | 20 (47.6%) | 12 (28.6%) | 3 (7.1%) | 6 (14.3%) | 2 (4.8%) | 4 (9.5%) | 14 (33.3%) | 16 (38.1%) |

introduced the theoretical foundation of the aggregation algorithm, explaining how it addresses disagreements across LIME, SHAP, and BreakDown. The second part focused on the practical evaluation of XMENTOR, our VS Code plugin. For this phase, participants were provided with installation requirements and step-by-step video guides to set up the plugin on both Windows [1] and Linux [2] systems. A pilot study with five practitioners was conducted to assess the clarity and length of the study. Based on their feedback, minor refinements were made. The final survey, estimated to take 25–30 minutes across both parts, was organized into six sections, with pilot responses excluded from the final dataset.

**(i) Consent and Prerequisite:** Participants confirmed their consent and familiarity with XAI techniques, AI/ML, and Python. Those without prior experience in XAI were not allowed to proceed.

**(ii) Participant Information:**This section gathers participants' details, including their Python experience, profession, and country.

**(iii) Disagreement Challenges:** Participants viewed examples of disagreements between explanation methods and answered questions on feature, sign, and rank disagreement, sharing which they found most severe and why.

**(iv) Impact of the Aggregation:** We ask participants to validate the novel method by confirming if the aggregation mechanism truly reduces perceived disagreement between explanations and whether they would prefer the tool XMENTOR in their daily workflows.

**(v) Participant's Recommendation:** We ask participants about the practical aspects of XMENTOR, including installation complexity, any technical issues encountered, system response time, and whether they would recommend the tool to other practitioners.

**(vi) Usefulness of XMENTOR:** We asked participants how useful they found the extension compared to not using it, what type of explanation they preferred like aggregated, individual, or side-by-side view, and what additional features they would like to see in XMENTOR to make the approach more robust and effective.

**Recruitment of Survey Participants:** Participants were recruited in two ways:

**(i) Snowball Approach:** We recruited participants worldwide through professional connections. Using a snowballing strategy [5], we encouraged initial participants to refer colleagues with similar experience and interest, creating a chain of qualified recommendations. In this approach, we found 30 participants. However, eight of them did not meet our requirements. We thus finally allowed the remaining 22 to participate in our survey.

[1] Windows
[2] Linux

**(ii) Open Circular:** To recruit participants, we shared our study details and research objectives in the targeted Facebook groups and LinkedIn for software developers. This approach resulted in 29 interested participants who met our criteria, ultimately yielding 20 valid responses.

Table 3 shows that most participants were Academicians (38.10%) or Software Engineers (33.33%), with about 80% having over three years of AI/ML experience. Figure 6 indicates that around 88% were familiar with XAI techniques. Participants represented diverse countries, ensuring the reliability and validity of the study's findings.

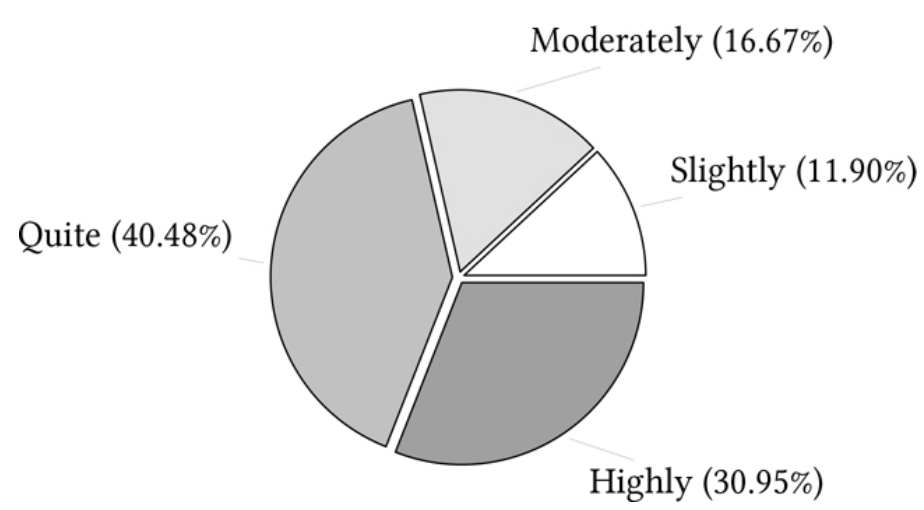

**Figure 6: XAI technique familiarity**

# 5 Findings & Discussion

## 5.1 Quantitative Findings: Disagreement Study

We applied LIME, SHAP, and BreakDown to multiple defect prediction datasets (e.g., ActiveMQ, Camel, Derby, Groovy, HBase, Hive, JRuby, Lucene, Wicket). Across these projects, the three techniques often produced inconsistent explanations.

**Extent of Disagreement:** Our analysis revealed that disagreements were common, especially when comparing LIME with BreakDown. SHAP and BreakDown as shown in Figure 8, in contrast, tended to align more closely.

**Types of Disagreement:** Among the three disagreement types: feature, rank, and sign, rank disagreements were the most common as shown in Figure 5, as explainers often agreed on important features but differed in their prioritization. Sign disagreements were less frequent but revealed cases where the same feature was assigned opposite effects by different methods.

**Implications:** The findings reveal that inconsistent explanations from different XAI methods can confuse developers and increase cognitive load, reducing trust and usability. This underscores the need for aggregation approaches like XMENTOR, which aim to harmonize explanations and ease the interpretability burden on practitioners.

## 5.2 User Study Outcomes

Our user study had two parts: a theoretical survey on disagreement concepts (feature, sign, rank) with an aggregation approach and a practical evaluation of the XMENTOR plugin. Participants could stop after the first part or continue with the second part, which required installing and using the tool inside VS Code. Among the total user study responses, five participants completed only the first part, while 37 installed and actively evaluated the plugin, giving us both conceptual feedback and practical assessments.

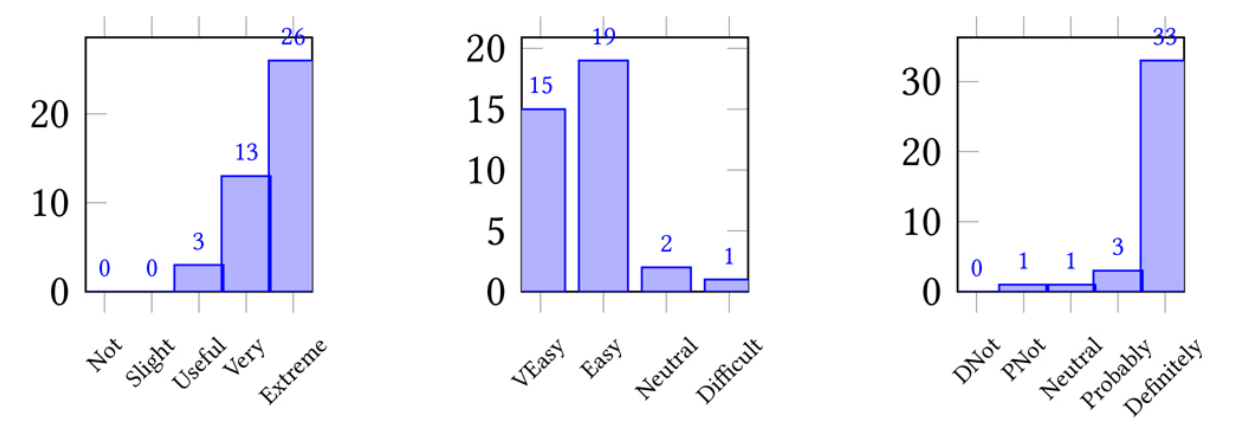

(a) Usefulness (b) Installation (c) Recommend

**Figure 7: Survey results: (a) usefulness of aggregation, (b) ease of installing XMENTOR, and (c) willingness to recommend.**

**Preferences:** Nearly 90% of participants expressed a clear preference for the aggregated explanation provided by XMENTOR. They described the unified view as more actionable and less confusing, noting that it allowed them to focus on the prediction rather than reconciling three conflicting outputs.

**Feedback:** The results were highly positive. As shown in Figure 7, aggregated explanations were consistently rated as valuable: nearly all participants marked them as very useful or extremely useful. Installation was straightforward; most rated it easy or very easy. Once installed, the tool performed smoothly, with fast response times. Most importantly, 33 participants (over 89.19%) said they would "definitely" recommend XMENTOR to others, indicating not just satisfaction but strong adoption potential.

Qualitative responses underscored these results. Several participants noted that rank disagreements were especially confusing: "*Some methods (like LIME) emphasize the local decision boundary for a single prediction, while others (like SHAP) balance both local and global signals. This mismatch leads to different orderings of feature importance.*". When asked about the plugin's usefulness, participants emphasized its impact on clarity and confidence: "*Strongly agree — having the extension makes explanations much clearer*" and "*With the aggregated view I don't waste time juggling three conflicting outputs.*" Others highlighted actionability, describing the aggregated explanation as "*very useful and easy to act on compared to separate explanations.*"

Together, these results demonstrate that practitioners not only understand the challenge of disagreement in theory but also recognize the practical value of resolving it through aggregation in an IDE plugin. By combining conceptual clarity with hands-on validation, our evaluation confirms that rank-aware aggregation makes explainability both more trustworthy and more usable in everyday developer workflows.

**Effectiveness:** Overall, participants reported that aggregated explanations not only improved clarity but also increased their trust in defect prediction. Rather than spending time and cognitive effort cross-checking different outputs, they could act directly on the unified explanation. This suggests that aggregation has the potential to reduce cognitive load, enhance usability, and make XAI tools more readily adoptable in real-world developer environments.

**Table 4: Average LLM Scores Across All Datasets**

| Method | Clarity ↑ | Completeness ↑ | Relevance ↑ |
|---|---|---|---|
| LIME | 3.65 | 3.15 | 3.59 |
| SHAP | 3.81 | 3.15 | 3.65 |
| BreakDown | 3.56 | 3.21 | 3.59 |
| **XMENTOR** | **3.75** | **3.28** | **3.68** |

Our study revealed clear contrasts in how developers experienced explanations with and without aggregation. When working with aggregated explanations, participants reported faster comprehension, less confusion, and higher trust. The unified view allowed them to focus on the most important features without the burden of reconciling multiple, and contradictory outputs. As one participant put it, "*Seeing one clear explanation is much better than juggling three conflicting ones.*" Another noted, "*The aggregated view removes confusion. I don't need to wonder which explainer to trust.*"

Interestingly, while the majority strongly preferred aggregation, a few participants pointed out that they sometimes valued the raw, unfiltered detail of individual explainers for debugging edge cases. As one remarked, "*For regular tasks the single view is better, but occasionally I'd still like to see the raw outputs in parallel.*" This suggests that while aggregation reduces cognitive load for most workflows, transparency and optional access to the underlying explanations remain important for advanced debugging scenarios.

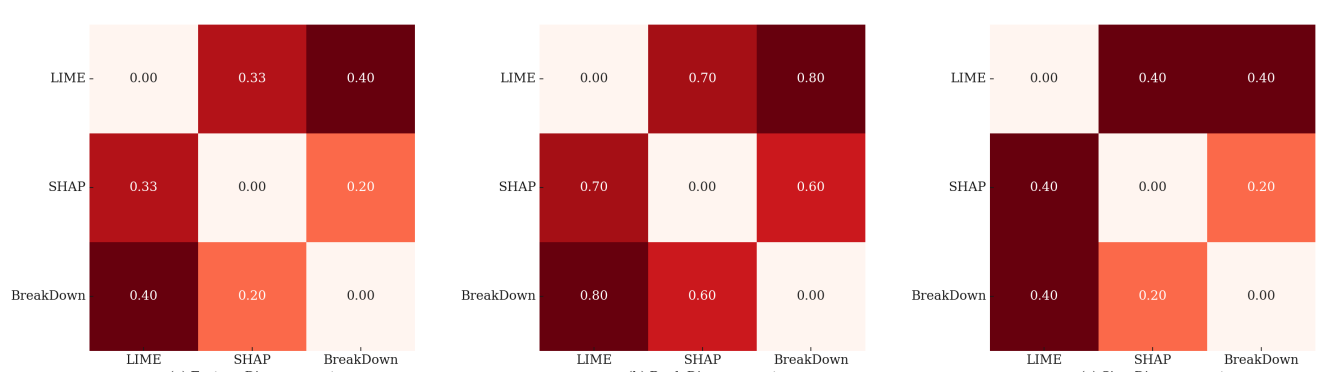

**Figure 8: Pairwise Disagreement Among XAI Techniques**

## 5.3 LLM-Assisted Explanation Quality

We evaluated explanations generated by LIME, SHAP, BreakDown, and XMENTOR using a Large Language Model (GPT-5) as an automated reviewer. For each explanation, the LLM assigns scores on a 1–5 scale along three dimensions like *clarity* (readability and understandability), *completeness* (coverage of key reasoning elements), and *relevance* (usefulness for a software developer's debugging), and provides a short natural-language justification for each score to support qualitative inspection alongside numeric comparison as shown in Figure 9. The evaluation covers 32 JIRA repository datasets, with LLM feedback revealing consistent explanation quality patterns across projects.

Table 4 shows that GPT-5 produces stable and discriminative judgments of explanation quality across datasets, clearly differentiating between LIME, SHAP, BreakDown, and XMENTOR along clarity, completeness, and relevance. While individual post-hoc methods achieve comparable scores, XMENTOR consistently attains the highest overall ratings, indicating that rank-aware aggregation improves perceived explanation quality by presenting clearer, more complete, and more developer-relevant explanations. At the same time, the LLM feedback repeatedly highlights a shared limitation across all methods: the lack of concrete remediation or next-step guidance for developers. Overall, these results support the use of LLMs as scalable and reliable proxies for expert reviewers when assessing explanation quality in software defect prediction.

**Answering RQ1: To what extent do popular post-hoc explanation methods (LIME, SHAP, BreakDown) disagree when applied to defect prediction tasks?**

Figure 8 shows that LIME and BreakDown disagree the most, whereas SHAP and BreakDown agree the most. We hypothesize that SHAP and BreakDown are most compatible as they both adhere to the additive property that features importance should be added to model predictions [22, 32]. Such a property is not satisfied by LIME, which could explain the higher disagreements. The additive property is that the contributions of all the features in an explanation should add up to the actual prediction of the model. Both SHAP and BreakDown follow this principle, ensuring their explanations are consistent with the model output. LIME, however, uses a simplified local approximation and doesn't enforce this property, which can lead to discrepancies between its attributions and the model's prediction. This difference likely explains why SHAP and BreakDown agree more, while LIME disagrees more with both.

We observe that LIME agrees slightly more with SHAP than with BreakDown. This may be because SHAP is a special case of LIME but uses a different proximity kernel [32]. However, without a ground truth, which explanation approach should they use? Aggregating all explanations may solve this difficulty. Before that, we examine the main disagreements of RQ2.

**Answering RQ2: Which type of disagreement, ranking inconsistencies or sign mismatches, is more prevalent among LIME, SHAP, and BreakDown?**

As stated previously, there are disagreements between explanations of files predicted as defects by ML models. However, some types of disagreements may occur more often than others. Indeed, once a set of features common to top-k of two explanations has been identified, two types of disagreements can occur between these features. Firstly, the features need not have the same ranks, which will result in a large value for $k(\text{FA} - \text{RA})$. Secondly, the features need not have the same sign, which will result in a great value for $k(\text{FA} - \text{SA})$. The histograms presented in Figure 10 suggests that disagreements in ranking occur far more often than disagreement in signs, among the top-k most essential features.

Indeed, $k(\text{FA} - \text{SA})$ are the most concentrated at the value zero. The fact that the distributions of $k(\text{FA} - \text{SA})$ are more spread rightward suggests that it is very uncommon for features common to top-k of two explanations to be given the same rank. Henceforth, one must be careful when making statements such as *this feature is more important than this one* because said statements may be supported by LIME but not by SHAP, or vice versa. Now that we have identified what type of disagreements occur most often, we study the possibility of aggregating explanations by focusing on the information on which they agree.

**Answering RQ3: Can a rank-aware aggregation method generate more consistent explanations by reconciling these disagreements?**

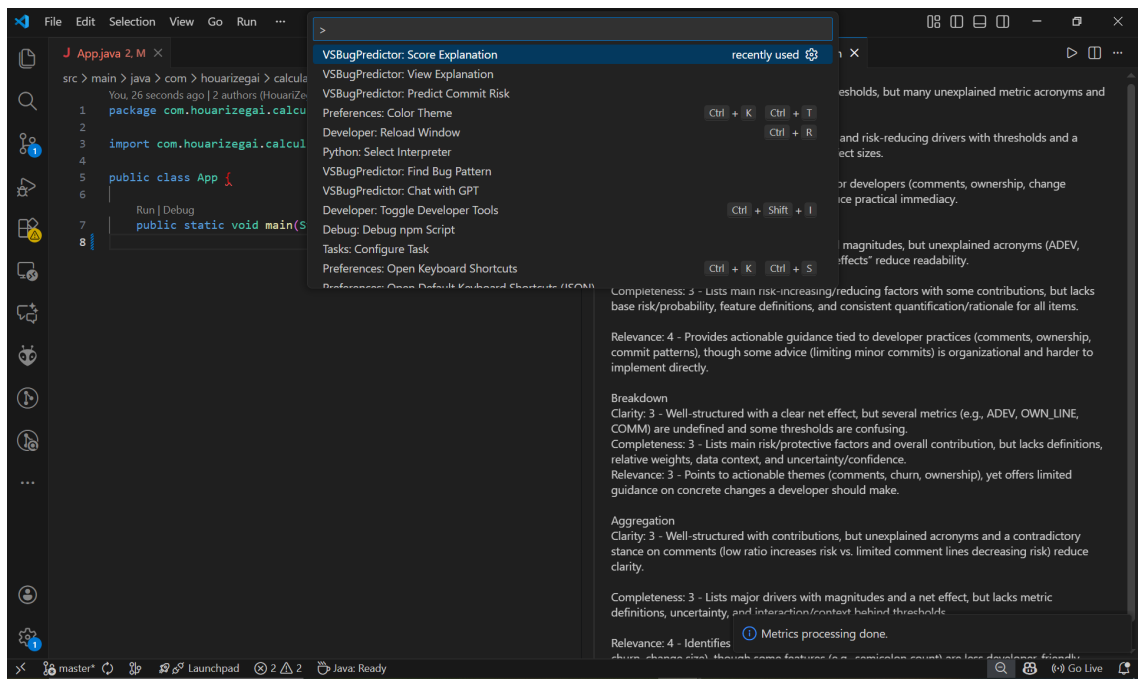

**(a) VS Code Command Palette Showing XMENTOR Prediction and Explanation Actions**

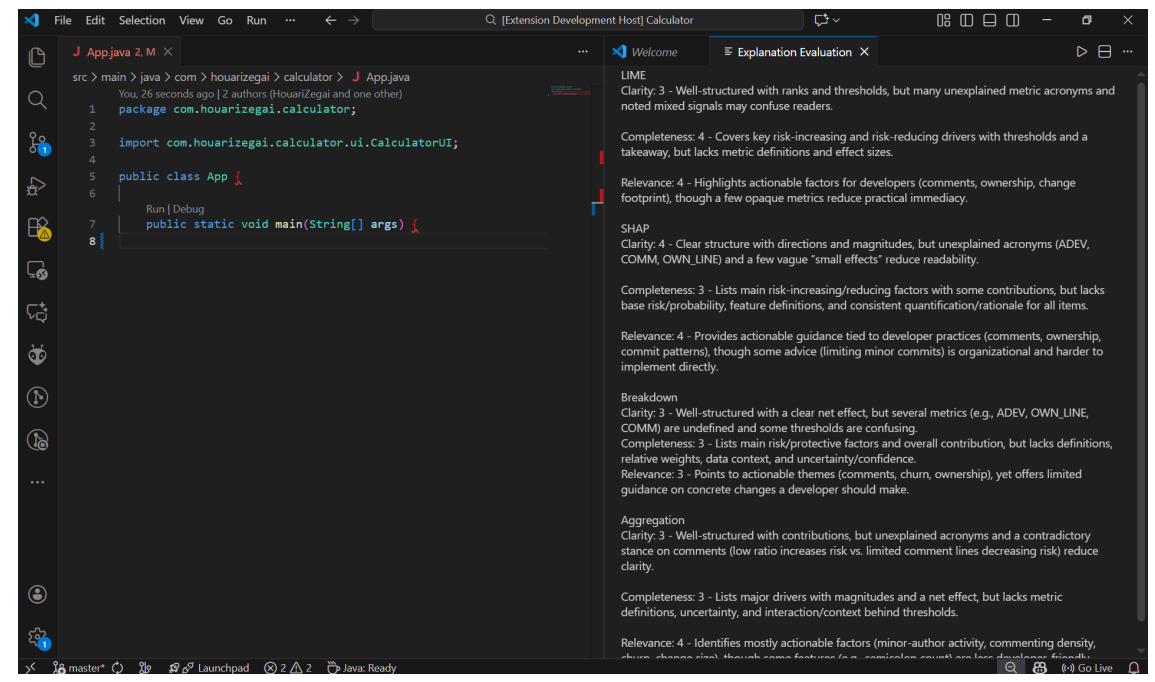

**(b) XMENTOR's Integrated Explanation Quality Feedback Inside the VS Code Environment**

**Figure 9: The LLM-assisted explanation evaluation panel has the scores for all the explanations and the comments for them.**

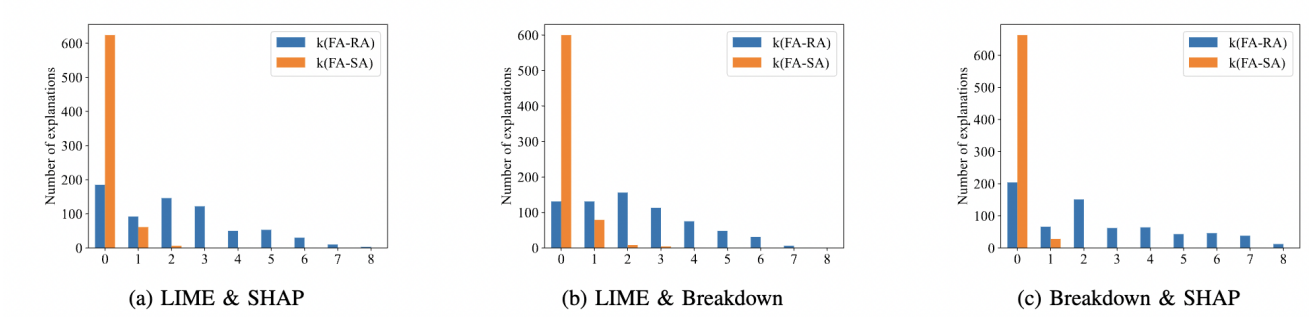

**Figure 10: Histograms of the differences between the FA metric and the RA and SA metrics for two explainers.**

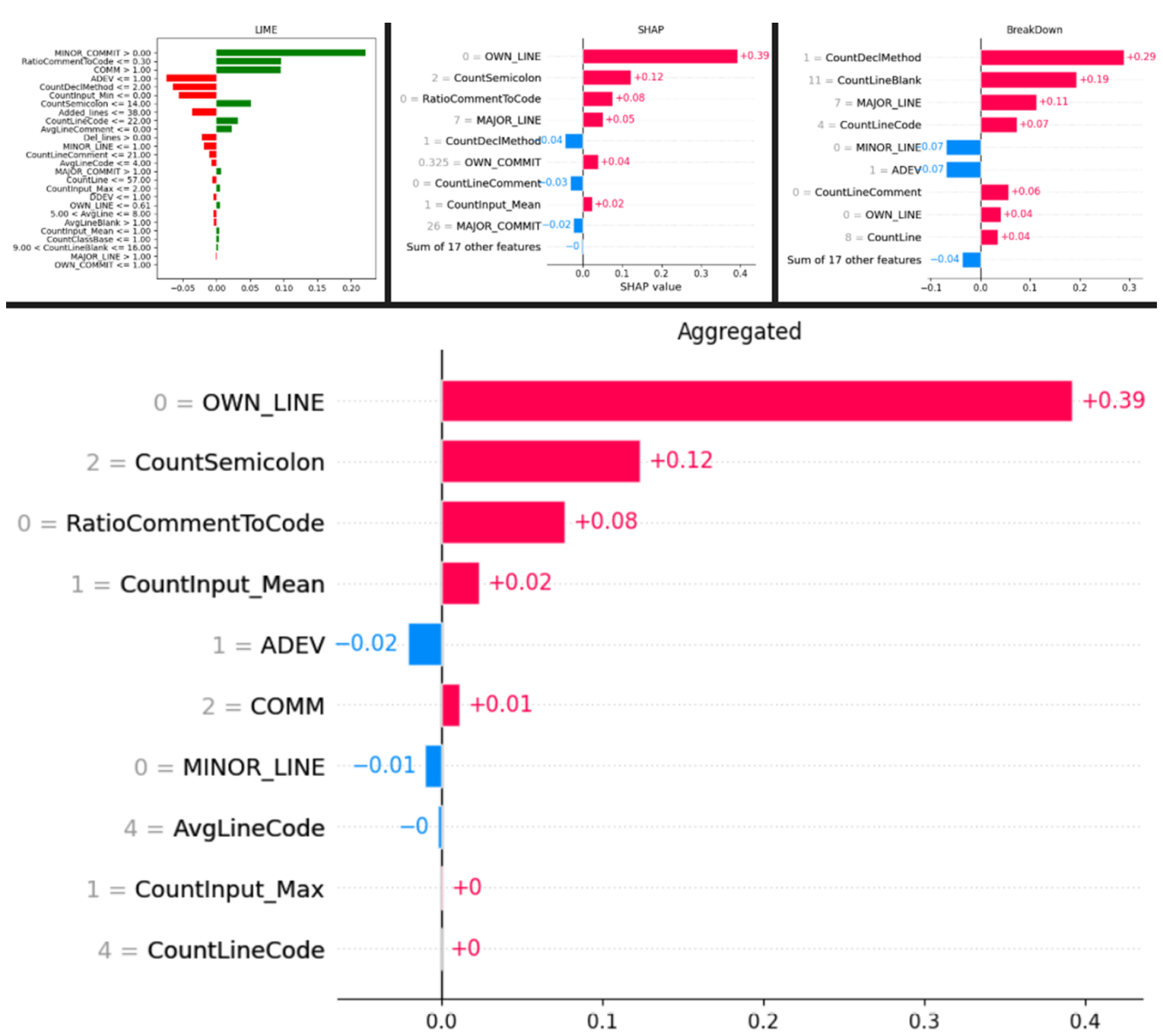

**Figure 11: Unified explanations: aggregation resolving disagreements across LIME, SHAP, and BreakDown**

A rank-aware aggregation method can generate more consistent explanations by reconciling disagreements across LIME, SHAP, and BreakDown. As shown in the Figure 11, each individual explainer highlights different features: LIME emphasizes churn-related metrics, SHAP prioritizes OWN_LINE and CountSemicolon, while BreakDown points to CountDeclMethod and CountLineBlank. When viewed separately, these outputs create a fragmented and sometimes contradictory picture of why the model flagged a file as buggy.

By applying our rank-aware aggregation approach as discussed in the subsection 3.3, these conflicts are resolved into a unified explanation. The aggregated view (bottom of the Figure 11) consistently highlights OWN_LINE, CountSemicolon, and RatioCommentToCode as the top contributors, with stable ranking and clear directional signs. Instead of forcing developers to reconcile three competing explanations, the aggregated output provides a single, coherent narrative that balances agreement across methods while ensuring coverage of important features.

This example demonstrates that aggregation does not simply average results but reduces confusion, lowers cognitive load, and presents explanations in a way that is both interpretable and actionable in practice.

**Answering RQ4: How do practitioners perceive the usefulness and clarity of aggregated explanations, particularly when integrated into a familiar development environment such as VS Code?**

Our user study revealed that practitioners found aggregated explanations both useful and clear, especially when integrated into the XMENTOR plugin. Most participants (85%) noted that rank inconsistency was the most problematic form of disagreement, underscoring the value of a rank-aware aggregation approach. One of the participants mentioned, "*Rank disagreements are severe in XAI because different methods capture different aspects of a complex model's behaviour.*" When evaluating the tool in practice, 35% rated XMENTOR as very useful and 59% as extremely useful for everyday tasks such as reviewing defect-prone files and debugging. Installation was also considered straightforward: approximately half said it was easy and nearly 40% said it was very easy, with 72.97% reporting no issues and the remainder encountering minor and moderate challenges. Performance was highly rated, with more than 90% describing the plugin's response time as fast or very fast. Nearly 89% stated they would recommend XMENTOR to other practitioners.

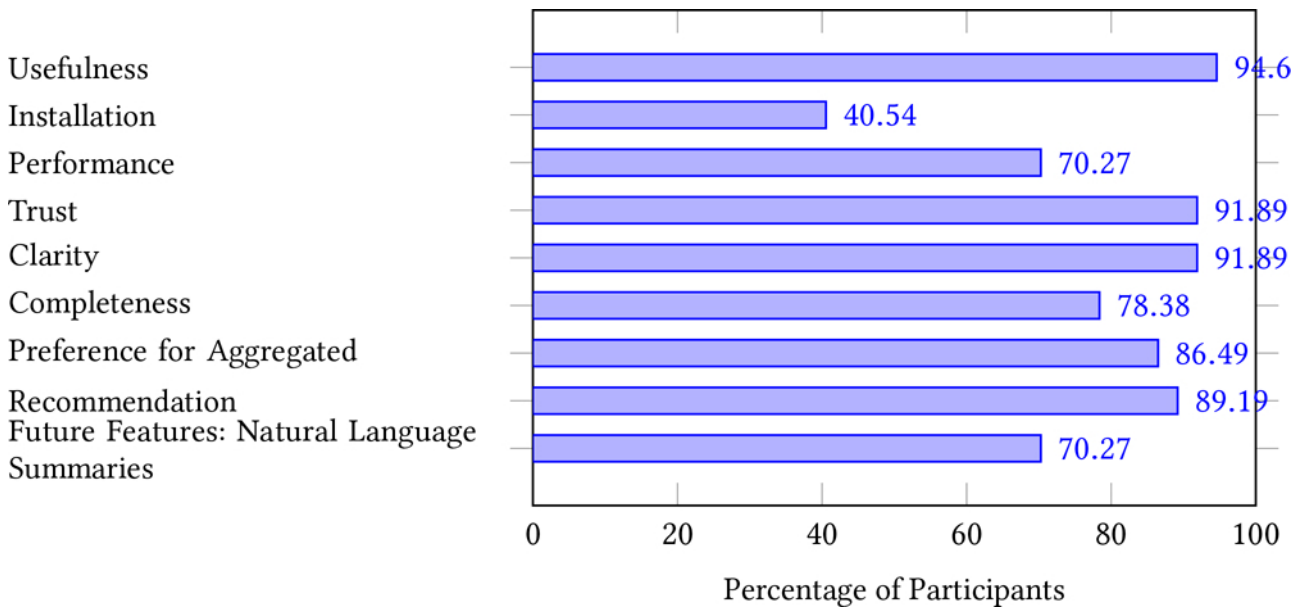

**Figure 12: Practitioner perceptions of XMENTOR aggregated explanations (RQ4).**

Figure 12 shows the practitioners' perceptions of XMENTOR on various dimensions.

Participants strongly preferred the aggregated explanation over individual ones: 86.49% chose aggregation as their default view, while 29.73% valued seeing both side by side. They emphasized that the aggregated explanation was easy to understand (59.46% strongly agreed), comprehensive (37.84% strongly agreed), and trust enhancing (91.89% agreed or strongly agreed). Every participant agreed that the tool was useful for decision-making, highlighting its impact compared to working without it. Looking ahead, 70.27% suggested incorporating natural language summaries to translate technical outputs into plain English, and 89.19% recommended adding Top-N Feature Rationales to explain why specific features mattered. These findings suggest that practitioners not only see aggregation as a solution to confusing, conflicting explanations but also view XMENTOR as a practical and trustworthy tool that fits naturally into their development workflow.

## 6 Implications and Broader Impact

Our findings suggest that XMENTOR has clear implications for both practitioners and the design of human-centered explainable AI systems. By reducing disagreements among LIME, SHAP, and BreakDown, XMENTOR demonstrates how aggregation can transform fragmented, and at times confusing, outputs into a coherent narrative that developers can act upon with confidence. This not only lowers cognitive load but also builds trust in defect prediction models, a factor often cited as a barrier to adoption in practice.

Beyond defect prediction, the design of XMENTOR highlights a more general approach, such as embedding rank-aware aggregation into everyday development tools, which can make explainability more usable and accessible. Developers benefit when explanations are integrated directly into their workflow rather than presented as external, abstract artifacts. The approach could be extended to other tasks where multiple explainers compete for attention such as code summarization, commit message classification, or bug localization and even to domains beyond SE where practitioners face similar challenges of reconciling conflicting model outputs.

At a broader level, XMENTOR contributes to trust, interpretability, and intelligibility in AI systems. Its design underscores the importance of not only providing explanations but also ensuring those explanations are consistent, actionable, and situated in context. By prioritizing agreement and usability, XMENTOR points toward a new class of human-centered explainability tools that move beyond simply "opening the black box" to delivering explanations that genuinely support decision-making.

## 7 Threats to Validity

**Internal Validity:** The primary concern is the performance of ML models due to dataset imbalance. We applied SMOTE [46] to address this. Additionally, ML pipelines involve stochastic elements such as train/test splits and hyperparameter tuning. To mitigate this, results were consolidated across multiple models. The SA metric was used as a proxy for measuring disagreement to yield promising preliminary findings. While we carefully designed the aggregation algorithm and user study, some factors may have influenced the results. For example, participants' prior familiarity with LIME, SHAP, or BreakDown could have shaped how they perceived individual versus aggregated explanations. To mitigate this, we recruited participants with varying levels of experience and provided background context before tasks.

**External Validity:** Our study focuses on within-project defect prediction, while other scenarios such as just-in-time [25, 34], cross-project [9, 18], or heterogeneous defect prediction [49] may yield different results. The dataset size may limit generalizability, though it was carefully selected for reliability. Future studies should explore broader datasets and assess generalization across project versions. Our findings are based on defect prediction datasets and practitioner feedback collected in controlled settings. While participants represented a range of backgrounds, they may not fully capture the diversity of software projects, organizational cultures, or toolchains in industry. As a result, generalizing beyond the studied tasks and IDE environment (VS Code) were done with caution.

**Construct Validity:** The study relies on feature, rank, and sign agreement metrics to quantify disagreement, which may not fully capture explainability inconsistencies, such as feature interactions [28]. Aggregation validation was based on practitioner perceptions rather than direct decision-making accuracy. While the study's defect prediction datasets cover diverse scenarios, findings may not generalize to all software contexts. Despite these limitations, the use of established metrics, and practitioner validation provides a solid foundation for further refinement.

## 8 Conclusion and Future Work

In this paper, we position XMENTOR as a solution to an important HCI challenge when post-hoc explainers such as LIME, SHAP, and BreakDown provide conflicting outputs, developers face confusion, frustration, and increased cognitive load. We addressed this usability problem by introducing a rank-aware aggregation approach that prioritizes agreement, producing explanations that are not only technically consistent but also more actionable for practitioners. Our findings show that rank disagreements dominate and that developers strongly prefer unified explanations integrated into their everyday environment. By embedding XMENTOR into VS Code, we demonstrate how explanations can become part of the workflow rather than an external burden. More broadly, this work highlights the need for human-centered XAI: explanations must be designed for usability and trust, not just accuracy. Future research should explore adaptive explanations, broader workflow integration, and large-scale studies with diverse developer communities.

## Acknowledgments

This research is supported in part by the Natural Sciences and Engineering Research Council of Canada (NSERC) Discovery Grants program, the Canada Foundation for Innovation's John R. Evans Leaders Fund (CFI-JELF), and by the industry-stream NSERC CREATE in Software Analytics Research (SOAR).